\documentclass[preprintnumbers,aps,prl,amsmath,amssymb,showpacs,twocolumn]{revtex4}
\begin{document}

\preprint{AEI--2004--099}

\title{Spherically Symmetric Quantum Horizons}

\author{Martin Bojowald}
\email{mabo@aei.mpg.de}
\affiliation{Max-Planck-Institut f\"ur Gravitationsphysik, Albert-Einstein-Institut,\\
Am M\"uhlenberg 1, D-14476 Potsdam, Germany}

\author{Rafal Swiderski}
\email{swiderski@aei.mpg.de}
\affiliation{Max-Planck-Institut f\"ur Gravitationsphysik, Albert-Einstein-Institut,\\
Am M\"uhlenberg 1, D-14476 Potsdam, Germany}


\newcommand{\lP}{\ell_{\mathrm P}}
\newcommand{\vp}{\varphi}
\newcommand{\vt}{\vartheta}

\newcommand{\md}{{\mathrm{d}}}
\newcommand{\Kern}{\mathop{\mathrm{ker}}}
\newcommand{\tr}{\mathop{\mathrm{tr}}}
\newcommand{\sgn}{\mathop{\mathrm{sgn}}}

\newcommand*{\R}{{\mathbb R}}
\newcommand*{\N}{{\mathbb N}}
\newcommand*{\Z}{{\mathbb Z}}
\newcommand*{\Q}{{\mathbb Q}}
\newcommand*{\C}{{\mathbb C}}

\begin{abstract}
  Isolated horizon conditions specialized to spherical symmetry can be
  imposed directly at the quantum level. This answers several
  questions concerning horizon degrees of freedom, which are seen to
  be related to orientation, and its fluctuations at the kinematical
  as well as dynamical level. In particular, in the absence of scalar
  or fermionic matter the horizon area is an approximate quantum
  observable. Including different kinds of matter fields allows to
  probe several aspects of the Hamiltonian constraint of quantum
  geometry that are important in inhomogeneous situations.
\end{abstract}

\maketitle

Black holes are the main objects, besides cosmology, where the
interface between classical gravity and expected properties of quantum
gravity is strongest. Not only are they classically singular, which
has to be resolved by quantum gravity \cite{BHPara}, but also their
horizons, which for massive black holes are far away from the strong
curvature region around the singularity, have provided several puzzles
that have been motivations for quantum gravity developments over
several decades. Most influential was the observation that an entropy
can be associated with a black hole horizon whose microscopic degrees
of freedom cannot be accounted for classically. This problem has been
solved, in the most general case of astrophysically relevant black
holes, by detailed calculations in quantum geometry
\cite{LoopEntro}. Isolating the relevant microscopic
degrees of freedom responsible for black hole entropy was possible
only by using new developments which replace the concept of an event
horizon by the quasilocal definition of an isolated horizon
\cite{HorRev}. A surface intersecting the isolated horizon was
then used as an inner spatial boundary which carries degrees of
freedom describing the horizon geometry that are matched to the bulk
quantum geometry. The matching is non-trivial and provides a
consistency test of the methods of quantum geometry.

There have been other expectations concerning horizons in addition to
the fact that they should carry microscopic degrees of freedom. Partly
motivated from possible explanations of the entropy, it has been
suggested that the horizon in quantum theory would not be a sharp
surface but would fluctuate. Also other aspects of quantum horizons,
which cannot be tested when the horizon is introduced as a boundary,
are of interest and necessary for a complete understanding of effects
such as Hawking radiation. For this reason, one would like to `quantize'
the isolated (or even dynamical) horizon conditions and impose them on
states at the quantum level. This is what we will do, in a first step,
in this article. Our analysis is complementary to that in
\cite{LoopEntro} in that we use the same ingredients ---
isolated horizons and quantum geometry --- but impose all the horizon
conditions at the quantum rather than some of them at the classical
level. Since this, at the current stage of developments, is too
complicated to do in the full theory, we do the analysis in spherical
symmetry. Even though in this case the full machinery of isolated
horizons would not seem to be necessary classically, we will see that
those conditions are important to decide how a horizon is to
be found at the quantum level. Despite the classical simplicity of
the Schwarzschild solution, we are able to learn much about the
quantum horizon structure such as the localization of the horizon,
its degrees of freedom, and its area as an observable.

\paragraph{Isolated horizon conditions.} There are three main parts to
the definition of an isolated horizon $\Delta$ with spatial sections
$S\cong S^2$ of given area $a_0$ \cite{IHPhase,ALRev}:

i) The canonical fields $(A_a^i,E^a_i)$ on the horizon are
 completely described by a single field $W=\frac{1}{2}\iota^*A^ir_i$
 on $S$ which is a U(1)-connection obtained from the pull-back of the
 Ashtekar connection to $S$. Here, $\iota\colon S\to\Sigma$ is the
 embedding of the horizon section $S$ into a spatial slice $\Sigma$
 and $r_i$ an internal direction on the horizon chosen such that $W$
 is a connection in the spin bundle on $S^2$ and $r^iE_i^a =
 \sqrt{\det q}\, r^a$ on the horizon with the internal metric $q$ on $S$
 and the outward normal $r^a$ to $S$ in $\Sigma$.

ii) The intrinsic horizon geometry, given by the pull-back of the
 2-form $\Sigma^i_{ab}:=\epsilon_{abc}E^c_i$ to $S$, is determined by
 the curvature $F=\md W$ of $W$ by
%
\begin{equation} \label{IsoCond}
 F=-\frac{2\pi}{a_0}\iota^*\Sigma^i r_i\,.
\end{equation}

iii) The constraints hold on $S$.

When the horizon is introduced as a boundary, condition i) is used to
identify the horizon degrees of freedom represented by the field
$W$. Condition ii) then shows that these degrees of freedom are fields
of a Chern--Simons theory on the horizon. It is the main condition
since it relates the horizon degrees of freedom to the bulk geometry,
which after quantization selects the relevant quantum states to be
counted. Condition iii), on the other hand, does not play a big role
since an isolated horizon as boundary implies a vanishing lapse
function on $S$ for the Hamiltonian constraint which then is to be
imposed only in the bulk.

When we impose the conditions at the quantum level it is clear that
the procedure will be very different. We will not be able to have an
independent boundary theory which is then matched to the bulk, but
would have to find the relevant degrees of freedom within the original
quantum theory. More importantly, we cannot use the simplification of
a vanishing lapse function since the horizon will no longer be
regarded as a boundary. In particular, the Hamiltonian constraint
will have to be imposed which in full generality is a daunting
task. For this reason we specialize the situation to spherical
symmetry which presents the simplest situation where horizons can
occur.

\paragraph{Spherical symmetry.} Spherically symmetric connections and
densitized vector fields are given by
%
\begin{eqnarray}
 A &=& A_x\tau_3\md x + A_{\vp} \Lambda_{\vt}^A \md\vt +
 A_{\vp}\Lambda_{\vp}^A \sin\vt\md\vp + \tau_3\cos\vt\md\vp\nonumber\\ 
E &=&  E^x\tau_3\sin\vt\partial_x + E^{\vp} \Lambda^{\vt}_E
 \sin\vt\partial_{\vt} + E^{\vp}\Lambda^{\vp}_E \partial_{\vp} \label{A}
\end{eqnarray}
in polar coordinates $(x,\vt,\vp)$. We use generators
$\tau_j=-\frac{i}{2}\sigma_j$ and choose a gauge in which the radial
components are along $\tau_3$. The angular components then are in the
$\tau_1$-$\tau_2$ plane and given by
$\Lambda_{\vp}^A=\cos\beta\tau_1+\sin\beta\tau_2$,
$\Lambda^{\vp}_E=\cos(\alpha+\beta)\tau_1+\sin(\alpha+\beta)\tau_2$,
and $\Lambda_{\vt}^A$ and $\Lambda^{\vt}_E$ given by rotating the
internal $\vp$-directions by ninety degrees around $\tau_3$ (see
\cite{SphSymm} for details). All fields $A_x$, $E^x$, $A_{\vp}$,
$E^{\vp}$, $\alpha$ and $\beta$ depend only on $x$, where $\beta$ is
pure gauge.

To evaluate the isolated horizon conditions, we
choose $r_i:=\sgn(E^x)\delta_{i,3}$ such that in fact $r^iE^a_i
= |E^x|\sin\vt\partial_x$ with the intrinsic horizon area element
$|E^x|\sin\vt$. Thus, $W=\frac{1}{2}r_i\iota^*A^i =
\frac{1}{2}\sgn(E^x) \cos\vt\md\vp$ whose integrated curvature given
by $\oint_S\md W = -2\pi\sgn(E^x)$ agrees with the Chern number of the
spin bundle, depending on the orientation given by $\sgn(E^x)$.

Evaluating (\ref{IsoCond}) first shows that in the spherically
symmetric context it is not restrictive since we have $a_0=4\pi |E^x|$
and the right hand side given by $-\frac{1}{2}\sgn(E^x)$ equals $F$
for all $E$. This is not surprising since the spherically symmetric
intrinsic geometry of $S$ is already given by the total area which is
fixed from the outset. Now the first condition plays a major role. A
further consequence of the isolated horizon conditions \cite{IHPhase}
is that the curvature ${\cal F}$ of the pull-back of $A_a^i$ to $S$ is
given by the curvature of $W$: $r_i{\cal F}(\iota^*A^i)=2\md
W$. Since ${\cal F}(\iota^*A)=
(A_{\vp}^2-1)\tau_3\sin\vt\md\vt\wedge\md\vp$, the condition requires
$A_{\vp}=0$ which will be the main restriction we have to impose on
quantum states in addition to the constraints.

This condition $A_{\vp}=0$ selects 2-spheres in a spherically
symmetric space-time corresponding to cross-sections of a
horizon. Indeed, for the Schwarzschild solution we have
$A_{\vp}=\Gamma_{\vp}$ since the extrinsic curvature
vanishes. Moreover, computing the spin connection for a spherically
symmetric co-triad $e_a$ yields the component
$\Gamma_{\vp}=e_{\vp}'/e_x$ which for the Schwarzschild
solution ($e_{\vp}=x$, $e_x=|1-2M/x|^{-1/2}$) yields the correct
condition $x=2M$.

\paragraph{Quantization.} Gauge invariant states of spherically
symmetric quantum geometry in the connection representation are given
by \cite{SphSymm}

\begin{equation} \label{GaugeInvSpinNetwork}
 T_{g,k,\mu}=\prod_e \exp\left(\tfrac{1}{2}i k_e
\smallint_e(A_x+\beta')\md x\right)  \prod_v
\exp(i\mu_v A_{\vp}(v))
\end{equation}
where $g$ is a graph in the radial manifold with edges $e$ and
vertices $v$ labeled by $k_e\in\Z$ and $\mu_v\in\R$. Connection
components act as multiplication operators, while spatial geometry is
encoded in the derivative operators

\vspace{-5mm}
\begin{eqnarray}
 \hat{E}^x(x) T_{g,k,\mu} &=& \frac{\gamma\lP^2}{8\pi}
\frac{k_{e^+(x)}+k_{e^-(x)}}{2} T_{g,k,\mu} \label{Exspec}\\
 \int_{\cal I}\hat{P}^{\vp}T_{g,k,\mu} &=& \frac{\gamma\lP^2}{4\pi}
\sum_{v\in{\cal I}} \mu_v T_{g,k,\mu}\label{Ppspec}
\end{eqnarray}
where the momentum $P^{\vp}=2E^{\vp}\cos\alpha$ of $A_{\vp}$
is integrated over intervals ${\cal I}$ in the radial manifold since
it is a density. Here, $e^+(v)$ and $e^-(v)$ are the edges neighboring
the vertex $v$ at larger and smaller $x$, respectively.

Geometrical operators can be obtained from the derivative
operators. The area operator \cite{AreaOp} for a sphere $S$ is simply
proportional to (\ref{Exspec}): $\hat{A}(S)=4\pi|\hat{E}^x(S)|$. The
volume operator is more complicated since the volume element depends
on $E^{\vp}$ which is a rather complicated function of $P^{\vp}$ and
$\alpha$. Nevertheless, it can be quantized and its full spectrum is
known \cite{SphSymmVol}. Just as in the full theory, its action is
non-zero only in vertices and its spectrum is discrete.

\paragraph{Quantum horizons.} We are now ready to impose the isolated
horizon condition on spherically symmetric states and to draw
conclusions for the quantum structure of horizons. As derived above,
the main condition is $A_{\vp}=0$ which can only be satisfied at a
vertex. If the condition is required strictly at a vertex $S$, the
$A_{\vp}$-dependence at $S$ must be a delta function
$\delta(A_{\vp})=\sum_{\mu}\exp(i\mu A_{\vp})$ which defines the
expansion in spin network states (\ref{GaugeInvSpinNetwork}). Such a
state is not normalizable in the kinematical inner product, and it is
not known what the situation would be for the physical inner
product. Fortunately, we can proceed without a detailed knowledge of
the state and just use the fact that a state having a quantum horizon
at a vertex $S$ is peaked on small values of $A_{\vp}(S)$. From
semiclassical considerations, which must hold at the horizons of large
black holes, it follows that $A_{\vp}=0$ cannot be imposed sharply, for
otherwise $P^{\vp}$ and the volume of regions around the horizon would
not behave classically.  The diffeomorphism constraint, which just
acts by moving the vertices along the radial manifold, can be averaged
as usually without changing the structure of the horizon vertex. We
will discuss the Hamiltonian constraint later.

It is immediate to see that fixing the horizon area, as usually done
in considerations of black hole horizon properties and their
thermodynamics, is consistent also at the kinematical quantum
level. Imposing the horizon condition just restricts the
$A_{\vp}$-dependence at the vertex, but leaves even the neighboring
edges and labels completely free. We can thus assume that our state is
an eigenstate of the area operator at $S$ and satisfies the horizon
condition there. Classically, this corresponds to the fact that
$A_{\vp}$ and $E^x$ have vanishing Poisson bracket.

The situation is different for the volume since it depends on
$P^{\vp}$. In fact, volume eigenstates have an $A_{\vp}$-dependence in
vertices given by Legendre functions \cite{SphSymmVol} which is
incompatible with the horizon condition. Thus, the volume of shells
around the horizon will not be sharp in quantum gravity. 

This observation allows to answer the question whether the horizon
will be a sharply localized surface. Looking at the state, the horizon
will be localized at a sharply defined vertex $S$, but this just
refers to localization in the background manifold. Moreover, after
solving the diffeomorphism constraint by group averaging the position
of the vertex will not be defined at all. As for physical
localization, we have to refer to a suitable measurement process. This
can easily be done by measuring the radial distance from an observer
in an asymptotic region at large $x$ to the horizon. The radial
distance is obtained by integrating the volume element divided by the
area element of spheres along the radial manifold, which is easily
quantized to an operator which has the same eigenstates as the volume
and area operator and acts as $\hat{L}(R)T_g = \sum_{v\in R\cap V(g)}
V_v/A_v T_g$ where the sum is over vertices of the graph $g$ in the
region $R$ between the horizon and the observer, and $V_v$ and $A_v$
are the volume and area eigenvalues, respectively, in a vertex $v$.

We now assume that the geometry in the asymptotic region up to regions
close to the horizon is semiclassical. Vertices outside the horizon
will then yield sharp contributions to the distance. But at the
horizon itself the volume cannot be sharply defined and thus the
location of the horizon itself as measured from outside is
unsharp, confirming older expectations of a fluctuating horizon
corresponding to a smeared-out region. Note
that this occurs in a way which is consistent with treatments
of the horizon as a sharp boundary as in entropy calculations. The
boundary refers to the background manifold according to which the
location is indeed sharp (at the boundary only tangential
diffeomorphisms generate gauge transformations).

While the structure of the horizon can well be analyzed in spherical
symmetry, the symmetry is too strong to preserve the microscopic
degrees of freedom. Detailed calculations of \cite{IHPhase} show that
horizon degrees of freedom are given by a U(1)-connection $W$, which in
spherical symmetry does not have free components. In fact, the
intrinsic geometry induced on a non-rotating isolated horizon as a
boundary by bulk quantum geometry is not spherically symmetric but
characterized by a finite set of punctures which endow the 2-sphere
with area. Thus, there are many more configurations describing a
non-rotating horizon than a spherically symmetric one. This has also
been indicated by attempts to derive black hole entropy from reduced
phase space quantizations, which have to introduce
degeneracies by additional arguments.

Here we see that there is only one binary degree of freedom to a
spherically symmetric isolated horizon, given by $\sgn(E^x)$. All
other components of the fields on the horizon are fixed either by the
required area $a_0$ or by the horizon condition $A_{\vp}=0$. The
situation remains the same in quantum theory: there are no new
quantum degrees of freedom. In particular, the reduction to spherical
symmetry removes almost all degrees of freedom counted in the black
hole entropy calculations of \cite{LoopEntro}. 

Still, there are surprising similarities to earlier considerations in
quantum gravity. First, there is one binary degree of freedom for an
exactly spherically symmetric horizon. If one imagines a non-spherical
horizon to be approximated as composed of spherically symmetric
patches this agrees with Wheeler's `It from Bit' picture
\cite{ItfromBit} (which has been generalized in the full calculation
\cite{Gamma}). Moreover, the binary degree of freedom is given
by $\sgn(E^x)=\sgn\det E^a_i$ which determines the orientation of
geometry at the horizon. This confirms the ideas of \cite{Orientation}
where the orientation of patches has been proposed to provide
gravitational horizon degrees of freedom. Note that these are indeed
{\em horizon degrees of freedom} since for an arbitrary surface the
pull back $A_{\vp}$ of $A$ would provide a further, continuous
parameter.

\paragraph{Dynamics.} We have seen that from the kinematical point of
view the horizon area even of a quantum black hole can be fixed
without contradicting the horizon conditions. However, the dynamical
point of view is more complicated since now, with the horizon not
being a boundary, the Hamiltonian constraint is non-trivial. We have
to check whether the area of an isolated horizon commutes with the
Hamiltonian constraint at least approximately. Even using
simplifications due to the symmetry the constraint is quite
lengthy, consisting of several terms. As in the full theory
\cite{QSDI} it is built from holonomies of $A$-components in (\ref{A})
some of which appear in commutators with the volume operator
$\hat{V}$.

Fortunately, the full expression simplifies under the isolated horizon
condition $A_{\vp}(S)=0$. Terms with $\sin A_{\vp}$, which appears in
holonomies, then annihilate states on which the horizon condition is
imposed sharply. As discussed above, the condition will not be sharp
in general, but still angular holonomies acting on a state can be
ignored compared to radial holonomies which depend on the unrestricted
$A_x$. Remaining terms are then of the form
\begin{equation} \label{H}
 \hat{H}\sim \sin(\tfrac{1}{2}\smallint A_x) \hat{V}
 \cos(\tfrac{1}{2}\smallint A_x)
 - \cos(\tfrac{1}{2}\smallint A_x) \hat{V}
 \sin(\tfrac{1}{2}\smallint A_x)
\end{equation}
where we wrote just one edge holonomy.

First, we can observe that the approximated constraint does not change
the $A_{\vp}$-dependence of a state and thus preserves the horizon
condition. With this result, we can then check if also the horizon
area is preserved, as expected from the classical vacuum behavior.
(Moreover, the Euclidean analysis of \cite{SphKl} shows that
$\sqrt{|E^x|}(1-A_{\vp}^2)$ is constant along the radial line and an
observable proportional to the ADM mass. At the horizon where
$A_{\vp}=0$ this specializes to $E^x$.)  Area is given by the labels
$k_e$ in (\ref{GaugeInvSpinNetwork}), which are changed by radial
holonomies appearing in (\ref{H}).  However, as in \cite{IsoCosmo} it
can easily be checked that they appear only in combinations which
leave the area eigenvalue invariant.  Thus, the area operator commutes
with (\ref{H}) which, taking into account that we ignored terms using
the horizon condition, implies that {\em the horizon area is an
  approximate quantum observable of spherically symmetric vacuum
  gravity}.  Even though expected classically, this result about the
{\em quantum observable} is non-trivial and depends on aspects of the
Hamiltonian constraint operator.  Using all terms in the constraint
shows that some of the neglected ones change the area, such that the
{\em quantum horizon area fluctuates dynamically}.

When a cosmological constant or an electromagnetic Hamiltonian
\cite{SymmRed} is added there are no new area-changing terms and the
result still holds true, again agreeing with classical
expectations. Coupling scalar or fermionic matter can
introduce terms which change the area eigenvalue such that here the
horizon can grow or shrink.

\paragraph{Conclusions.} At first sight, there apparently are several
possibilities to implement horizon conditions in spherically symmetric
situations. For instance, one can try to quantize $x=2M$ by using the
area operator for $x$ and the ADM mass for $M$. The drawbacks are that
this requires a mass operator (which, for instance, would be sensitive
to asymptotic flatness conditions) and that one part of the condition
would be quantized at the horizon, the other at infinity. Thus, one
would need a complete solution to the constraint in order to connect
both parts of the condition. (Expressions for the horizon mass
provided by the isolated horizon framework would work locally but make
the condition an identity.) Moreover, this procedure would not work
with matter.

The isolated horizon framework provides an unambiguous condition which
is local at the horizon. This makes it possible to impose the
condition without full knowledge of physical solutions, which to our
knowledge results in the first {\em implementation of horizon conditions
fully at the quantum level}. It is this isolated horizon
condition which leads to strong simplifications in the quantum
Hamiltonian constraint exploited here.

Our results verify some of the earlier expectations concerning
fluctuating horizons and make them more detailed. Moreover, we can
show that the horizon area is an approximate quantum observable in the
sense that it commutes with the dominant contribution to the
Hamiltonian constraint. These calculations test several aspects of the
constraint operator, in particular those which did not play a role in
homogeneous models \cite{Sing,IsoCosmo,Spin}. As we have seen,
going to the horizon simplifies the analysis of some aspects of
quantum observables since a horizon is much easier to impose on
quantum states than an asymptotic regime where one could test the ADM
mass.

The framework introduced here allows, e.g., to answer questions
related to black hole evaporation \cite{BHPara}. There are several new
possibilities not yet studied when matter Hamiltonians are coupled:
First, the horizon conditions need to be generalized to dynamical
horizons \cite{DynHor}, and whether or not the Hamiltonian
constraint will again simplify at the horizon depends on the precise
form of the conditions. A detailed analysis of the general situation
is yet to be undertaken, but at least for slowly evolving horizons
\cite{SlowHor} $A_{\vp}$ will be small: a horizon slowly evolving at a
rate $\epsilon$ (as defined in \cite{SlowHor}) has
$A_{\vp}\sim\epsilon$. Similar simplifications as used here will then
remain to hold true approximately in the slowly evolving case which
opens the prospect to investigate how quantum horizons grow when
matter falls in and shrink from Hawking radiation.

\smallskip

We thank Abhay Ashtekar, Hans Kastrup and Thomas Thiemann for
discussions.

\end{document}